\documentclass[final]{cvpr}

\usepackage{times}
\usepackage{epsfig}
\usepackage{graphicx}
\usepackage{amsmath}
\usepackage{amssymb}
\usepackage{algorithm}  
\usepackage{algorithmic} 
\usepackage{subfig}
\usepackage[table]{xcolor} 
\usepackage{multirow}
\usepackage{booktabs}
\usepackage{enumitem}

\newcommand{\Cmat}[0]{\ensuremath{{\bf C}} }
\newcommand{\Dmat}[0]{\ensuremath{{\bf D}} }
\newcommand{\Emat}[0]{\ensuremath{{\bf E}} }

\newcommand{\Gmat}[0]{\ensuremath{{\bf G}} }
\newcommand{\Hmat}[0]{\ensuremath{{\bf H}} }

\newcommand{\Omat}[0]{\ensuremath{{\bf O}} }

\newcommand{\Qmat}[0]{\ensuremath{{\bf Q}} }

\newcommand{\Wmat}[0]{\ensuremath{{\bf W}} }
\newcommand{\Xmat}[0]{\ensuremath{{\bf X}} }
\newcommand{\Ymat}[0]{\ensuremath{{\bf Y}} }
\newcommand{\Zmat}[0]{\ensuremath{{\bf Z}} }

\newcommand{\xv}[0]{\ensuremath{\boldsymbol{x}} }
\newcommand{\yv}[0]{\ensuremath{\boldsymbol{y}} }
\newcommand{\zv}[0]{\ensuremath{\boldsymbol{z}} }

\newcommand{\Gammamat}[0]{\ensuremath{\boldsymbol{\Gamma}} }

\newcommand{\Thetamat}[0]{\ensuremath{\boldsymbol{\Theta}} }

\newcommand{\alphav}[0]{\ensuremath{\boldsymbol{\alpha}} }
\newcommand{\betav}[0]{\ensuremath{\boldsymbol{\beta}} }

\newcommand{\mc}{\multicolumn}
\newcommand{\mr}{\multirow}

\newcommand\blfootnote[1]{%
	\begingroup
	\renewcommand\thefootnote{}\footnote{#1}%
	\addtocounter{footnote}{-1}%
	\endgroup
}

\usepackage[pagebackref=true,breaklinks=true,colorlinks,bookmarks=false]{hyperref}

\def\cvprPaperID{7697} 
\def\confYear{CVPR 2021}

\begin{document}

\title{MetaSCI: Scalable and Adaptive Reconstruction for Video Compressive Sensing}

\author{Zhengjue Wang$^{\dagger}$, Hao Zhang$^{\dagger}$, Ziheng Cheng, Bo Chen*\\
National Laboratory of Radar Signal Processing, Xidian University, Xian, China\\
{\tt\small \{zhengjuewang,zhanghao\_xidian\}@163.com,zhcheng@stu.xidian.edu.cn,bchen@mail.xidian.edu.cn}
\and
Xin Yuan*\\
Bell Labs,
NJ USA\\
{\tt\small xyuan@bell-labs.com}
}

\maketitle

\begin{abstract}
To capture high-speed videos using a two-dimensional detector, video snapshot compressive imaging (SCI) is a promising system, where the video frames are coded by different masks and then compressed to a snapshot measurement.
Following this, efficient algorithms are desired to reconstruct the high-speed frames, where the state-of-the-art results are achieved by deep learning networks.
However, these networks are usually trained for specific small-scale masks and often have high demands of training time and GPU memory, which are hence {\bf \em not flexible} to $i$) a new mask with the same size and $ii$) a larger-scale mask.
We address these challenges by developing a Meta Modulated Convolutional Network for SCI reconstruction, dubbed MetaSCI. 
MetaSCI is composed of a shared backbone for different masks, and light-weight meta-modulation parameters to evolve to different modulation parameters for each mask, thus having the properties of {\bf \em fast adaptation} to new masks (or systems) and ready to {\bf \em scale to large data}.
Extensive simulation and real data results demonstrate the superior performance of our proposed approach.  Our code is available at {\small\url{https://github.com/xyvirtualgroup/MetaSCI-CVPR2021}}.

\end{abstract}


\blfootnote{$^{\dagger}$Equal contributions. 
\quad  * Corresponding authors.
}

\vspace{-4mm}
\section{Introduction}

\begin{figure}[htbp!]
  \centering
  \subfloat[Adaptation to a new mask]{\includegraphics[width=.84\linewidth]{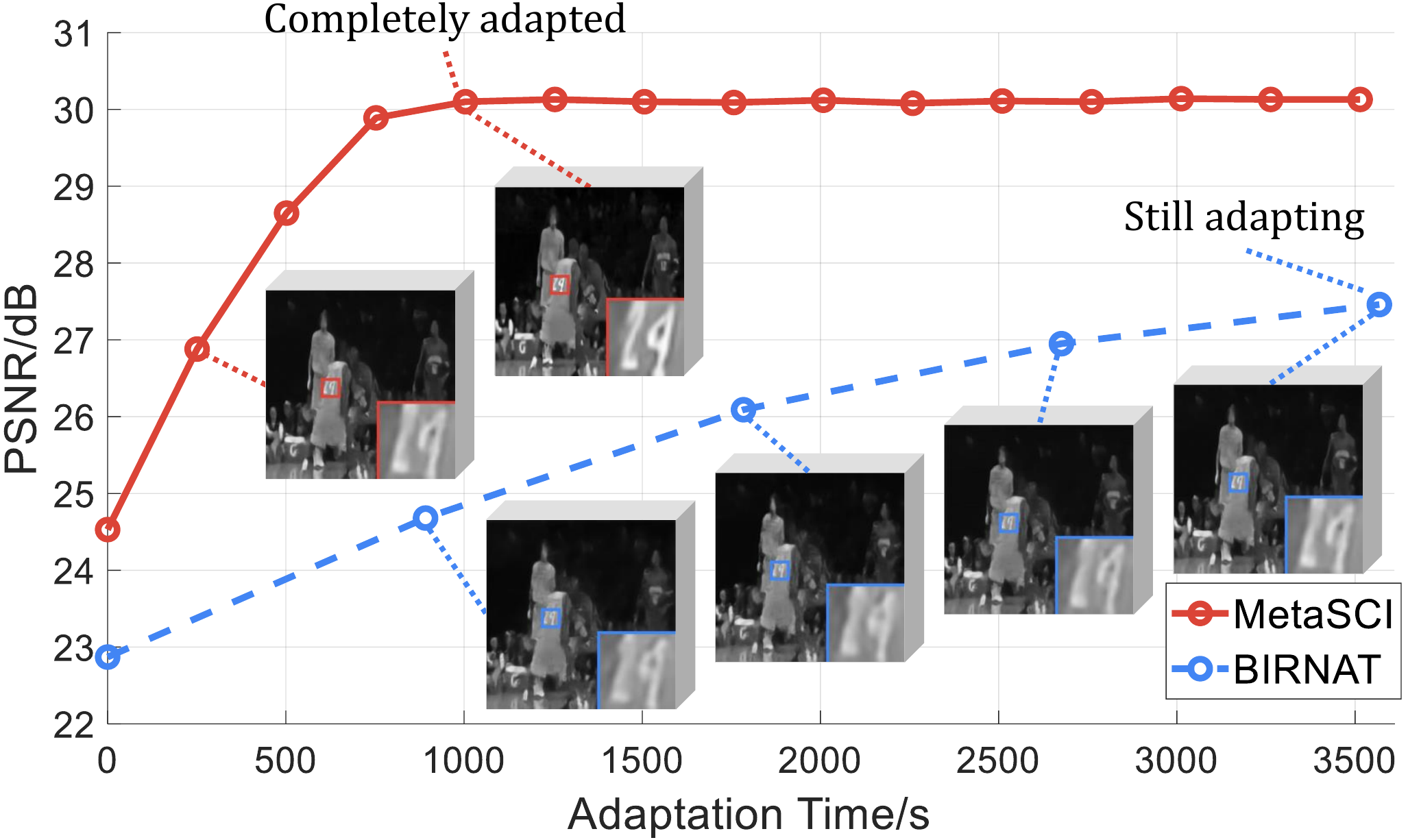}\label{fig1a}} \\
  \vspace{-2mm}
  \subfloat[MetaSCI for large-scale SCI reconstruction by fast adaptation]{\includegraphics[width=.84\linewidth]{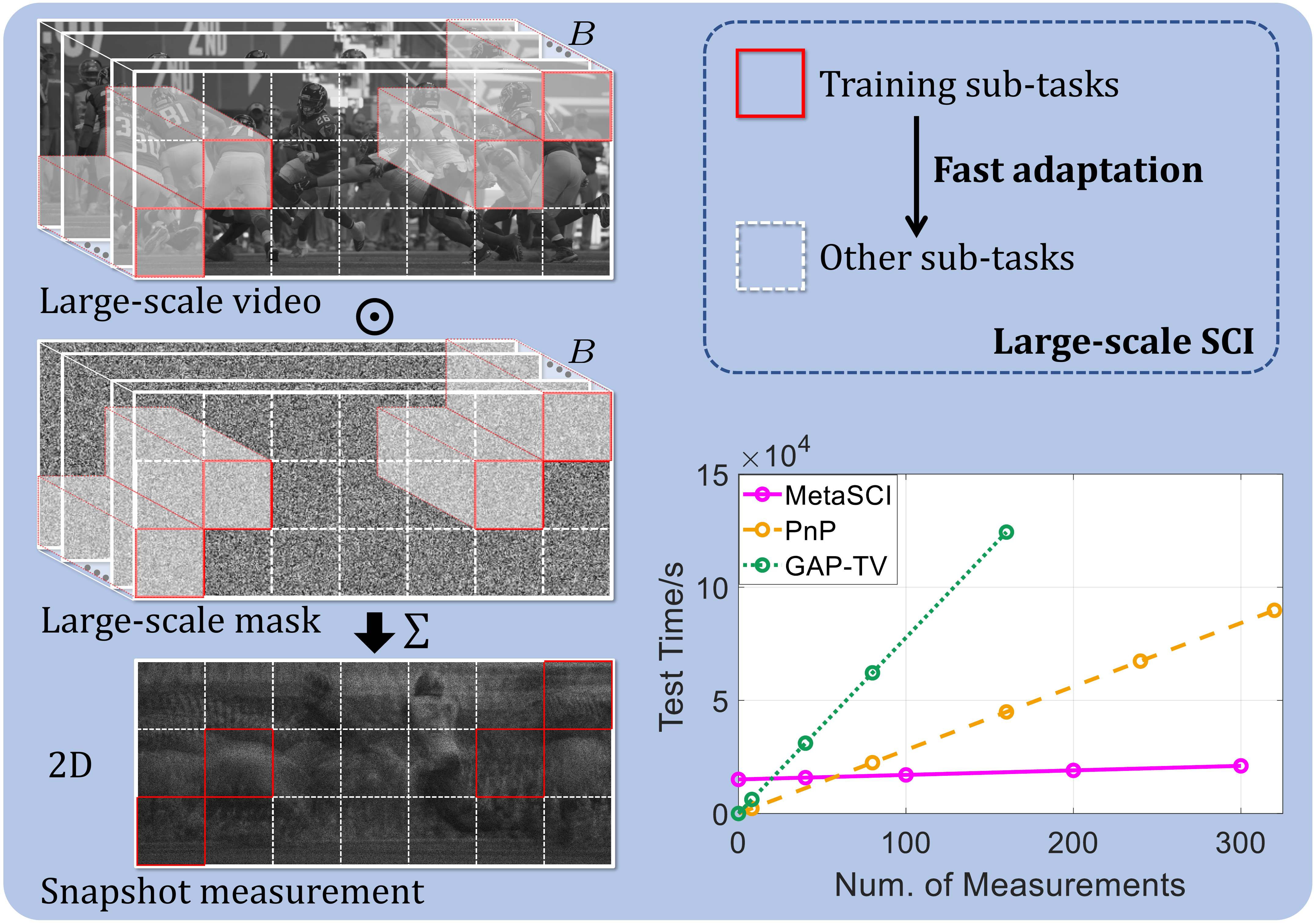}\label{fig1b}} 
  \vspace{-2mm}
  \caption{\small Illustration of the {\em fast adaptation} property of MetaSCI, as training on $256\times256$ measurements and adapted to (a) a new $256\times256$ measurement compressed by {\em different masks}, and (b) a $768\times 1792$ measurement compressed by large masks.
  The results in (a) is evaluated on a benchmark data, \textit{Kobe} \cite{liu2018rank}.
  Compared with BIRNAT \cite{cheng2020birnat}, a SOTA deep model, MetaSCI realizes much faster adaptation. 
  In (b), we decompose the large-scale SCI reconstruction task into 21 sub-tasks without overlap. MetaSCI is trained on 4 sub-tasks, and then fast adapted to the others.
  After the adaptation stage, MetaSCI can realize real-time reconstruction using feed-forward mapping.
  By contrast, PnP \cite{yuan2020plug} and GAP-TV \cite{yuan2016generalized}, the only two existing methods suitable for large-scale SCI, need iterative optimization for every measurement.
  }
  \label{fig1}
  \vspace{-7mm}
\end{figure}

High-speed video imaging system is desirable in our daily life, which often faces challenges in capturing and saving high-dimensional (HD) data, \eg, high memory, bandwidth and power demand.
Inspired by compressive sensing (CS)~\cite{Candes06_Robust,donoho2006compressed} techniques, video snapshot compressive imaging (SCI) has attracted much attention, which enjoys the advantages of low memory, low bandwidth, low power and potentially low cost \cite{yuan2020plug}.
The video SCI system constructs a pipeline of {an optical hardware encoder and a software decoder}~\cite{Yuan2021_SPM}.
In one exposure time, the optical encoder modulates the HD data via dynamic masks and then compresses  multiple high-speed frames into a two-dimensional (2D) snapshot measurement.
The decoder, on the other hand, aims to recover or reconstruct the high-speed video frames using advanced algorithms.
This paper focuses on video SCI reconstruction. 
More specifically, we develop a \textit{fast adaptive} decoder motivated by {\em meta learning}, which is flexible to different systems and ready to scale to large data.


In general, a desirable decoder should have good properties in $i$) high fidelity (often with PSNR $\geq$30dB) and $ii$) fast recovery.
With more than a decade of development, more and more optical encoders are constructed~\cite{hitomi2011video,llull2013coded,Qiao2020_CACTI,Qiao2021_MicroCACTI,qiao2020deep,reddy2011p2c2,Sun16OE,Sun17OE,Tsai15OL,yuan2014low,Yuan16BOE}, which arouses more considerations for a practical algorithm.
As a new encoding system is built, one may often wonder whether a well learned decoder can be \textit{fast adapted} to this new encoder.
A relatively simple scenario is that the physical masks are changed but with the same spatial size such as $256\times256$ pixels.
Even more challenging, the physical masks are scaled to higher spatial dimension such as $512\times512$ or even up to $2048\times2048$ pixels when a high resolution system is built.
Therefore, the properties of $iii$) \textit{fast adaptation} and $iv$) \textit{scalability} are also desired to make video SCI system being practical. However, existing SCI reconstruction networks often lack fast adaptation, \ie, not flexible. As the mask changes, the network has to be re-trained, which again needs a long time, as shown in Fig.~\ref{fig1a}. 

Bearing all these four-aspect concerns in mind, we propose a {\bf meta modulated convolutional network} for {\bf flexible SCI} reconstruction, dubbed MetaSCI.
With the following contributions and appealing properties, MetaSCI will pave the way of applying deep learning methods to large-scale SCI in our daily life.
\vspace{-2mm}
\begin{itemize}[leftmargin=10pt]
	\setlength{\itemsep}{0.1pt}
	\setlength{\parsep}{0pt}
	\setlength{\parskip}{2pt}

    \item We discuss the fast adaptation problem of video SCI in real applications, which have not been studied before, especially for deep learning models.  
	\item To realize fast adaptation, we construct a multi-encoding-system training regime and build MetaSCI. MetaSCI consists of a shared backbone for different systems, and light-weight meta-modulation parameters that can evolve to different modulation parameters for each individual system. 
	\item A hybrid learning algorithm is developed to train the network, with standard gradient descend for the shared backbone and meta updates for meta parameters. 
	\item Besides achieving competing performance and fast adaptation on widely used small-scale video benchmarks, using the attractive property of fast adaptation, MetaSCI is the first deep model to perform real-time large-scale video reconstruction as shown in Fig.~\ref{fig1b}.
\end{itemize}

\section{Related Work}
Existing SCI reconstruction models can be divided into two categories, optimization based ones and deep learning based ones.

The optimization based methods consider the ill-posed SCI reconstruction task as a regularized optimization problem, and usually solve it via iterative algorithms.
Among traditional methods \cite{bioucas2007new,liu2018rank,yang2014compressive,yang2014video,yuan2016generalized,Yang20_TIP_SeSCI}, DeSCI in \cite{liu2018rank} achieves the state-of-the-art (SOTA) reconstruction performance.
Yet, it takes more than one hour to recover a $256 \times 256 \times 8$ video, which precludes DeSCI to be applied to a higher-dimensional scenario.
To accelerate the optimization speed, Yuan \etal. \cite{yuan2020plug,PnP_SCI_arxiv2021} develop a plug-and-play (PnP) framework.
By plugging a pre-trained denoising network into every iteration of the optimization, PnP achieves comparable PSNR scores and much faster inference speed.
Generally, the major drawback of optimization based models is that one needs to perform iterative optimization for every new-coming snapshot measurement (sometimes needs to fine-tune the parameters), making it time-consuming as processing large amounts of snapshots, as shown in Fig.~\ref{fig1b}.

Compared with optimization based methods, the most attractive property of deep learning based ones is the real-time test speed.
In \cite{Cheng2021_CVPR_ReverSCI,cheng2020birnat,iliadis2018deep,iliadis2020deepbinarymask,ma2019deep,Meng_GAPnet_arxiv2020,meng2020end,meng2020snapshot,qiao2020deep}, various end-to-end networks are proposed, whose input is the measurement (and optionally masks) and the output is the recovered video, optimized by reconstruction loss such as mean square error (MSE) and adversarial loss \cite{goodfellow2014generative}.
Among them, the BIRNAT in \cite{cheng2020birnat} achieves comparable or even superior PSNR than DeSCI.
Even though, based on the bidirectional RNN structure, BIRNAT needs a significant amount of GPU memories, scarcely possible for large-scale SCI.
More importantly, as the mask changes, the network has to be re-trained, as shown in Fig.~\ref{fig1a}. 
Unfortunately, the training time is on the order of days or weeks.

\begin{table}
    \centering
        \caption{\small Property of typical SCI reconstruction algorithms.}
        \vspace{-3mm}
    \resizebox{0.48\textwidth}{!}{
    \begin{tabular}{c|c|c|c|c}
        \toprule
        Algorithm & Fidelity & Inference speed & Fast adaptation & Scalability  \\
        \midrule
        DeSCI \cite{liu2018rank}     & High     & Low             & Middle             & Low  \\  
        PnP \cite{yuan2020plug}       & High     & Middle          & Middle         & High  \\
        BIRNAT \cite{cheng2020birnat}    & High     & High            & Low             & Low  \\
         \rowcolor{lightgray}
        MetaSCI   & High     & High            & High            & High  \\
        \bottomrule
    \end{tabular}}
    \label{tab:related work}
    \vspace{-3mm}
\end{table}

To make more evident comparisons, Table~\ref{tab:related work} summarizes the property of some typical SCI reconstruction algorithms\footnote{Note that though DeSCI and PnP can easily adapt to new systems, one usually needs to fine-tune the parameters to achieve good results. Therefore, we dub their adaptation to `middle'.}, highlighting our motivations to propose a scalable and adaptive reconstruction model for video SCI.

\begin{figure*}[!t]
    \centering
    \includegraphics[width=1\textwidth]{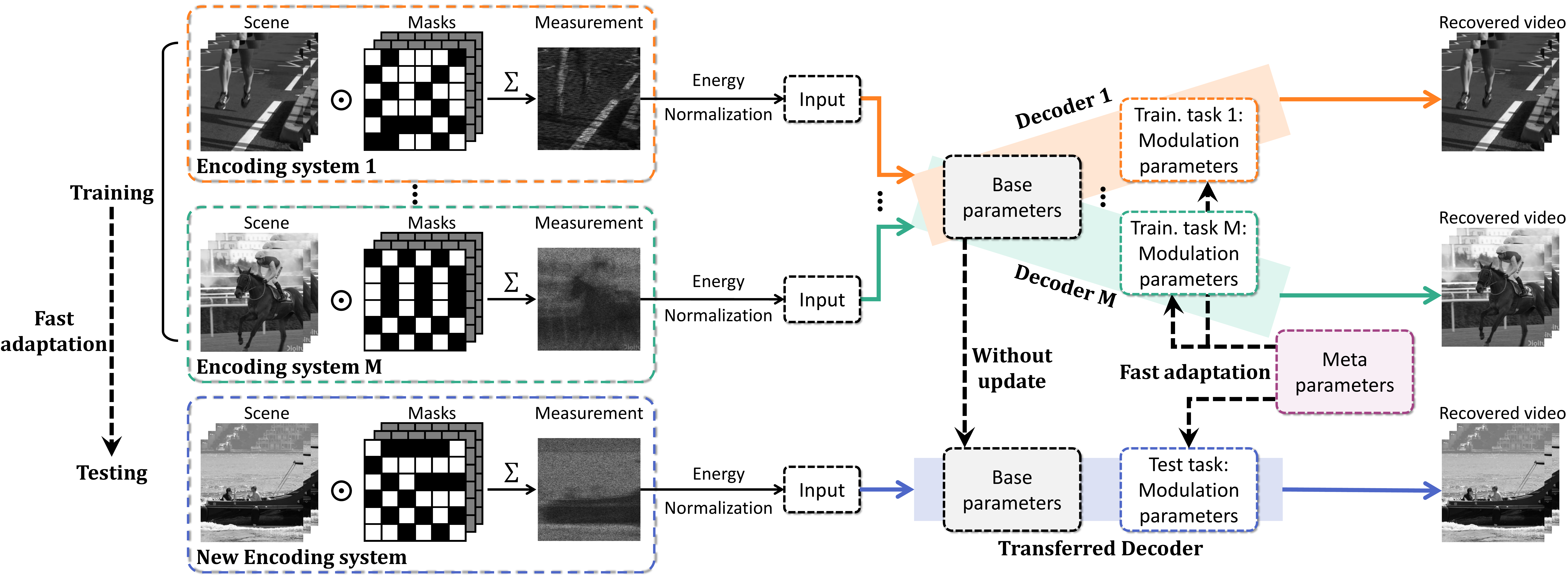}
    \caption{Pipelines of video SCI (\textit{optical encoder}) and our proposed MetaSCI (\textit{software decoder}) for reconstruction. A video scene, represented by a sequence of images, is coded ($\odot$) by binary masks and then integrated ($\sum$) over time on a camera, forming a single-frame compressed measurement. At the training stage, by minimizing MSE between recovered and original videos, we use $M$ masks (tasks) to train our proposed network including shared base parameters and meta parameters, where meta parameters evolve to modulation parameters, modulating (see Fig. \ref{fig2: model_2} for details) the shared base parameters for respective tasks. After training, we get well-learned base and meta parameters. During testing, we have a new mask, fixing the base parameters, we only need to run a few iterations to update well-learned meta parameters to fit our new mask, realizing fast adaptation.}
    \label{fig: model_1}
    \vspace{-3mm}
\end{figure*}

\section{Problem Statement}\label{sec: SCI and adaptation}
We first provide the mathematical model of video SCI, and then discuss the importance of fast adaptation in SCI.

\subsection{Mathematical Model of Video SCI}\label{sec: SCI}
We assume that a scene with $B$ high-speed frames $\{\Xmat_b\}_{b=1}^B \in \mathbb{R}^{d_x \times d_y}$ is modulated by the coding patterns (masks) $\{\Cmat_b\}_{b=1}^B \in \mathbb{R}^{d_x \times d_y}$.
As shown in Fig. \ref{fig: model_1}, a measurement $\Ymat \in \mathbb{R}^{d_x \times d_y}$ is then obtained by
\begin{equation}\label{eq: sci1}
\textstyle \Ymat = \sum_{b=1}^B \Xmat_b \odot \Cmat_b + \Zmat,
\end{equation}
where $\odot$ and $\Zmat \in \mathbb{R}^{d_x \times d_y}$  denote the matrix element-wise product and noise, respectively.
Denoting the vectorization operation on a matrix as $\mbox{Vec}(\cdot)$, \eqref{eq: sci1} can be re-written as
\begin{equation}\label{eq: sci2}
\yv = \Hmat \xv + \zv,
\end{equation}
where $\yv = \mbox{Vec}(\Ymat) \in \mathbb{R}^{d_x d_y}$ and $\zv = \mbox{Vec}(\Zmat) \in \mathbb{R}^{d_x d_y}$.
Correspondingly, the video $\xv$ is expressed as
\begin{equation}\label{eq: sci3}
\xv = \textstyle \left[ \mbox{Vec}(\Xmat_1)^T, \cdots, \mbox{Vec}(\Xmat_B)^T \right]^T \in \mathbb{R}^{d_x d_y B}.
\end{equation}
Unlike traditional CS methods \cite{donoho2006compressed,needell2010signal} using a dense sensing matrix, in SCI, the sensing matrix $\Hmat \in \mathbb{R}^{d_x d_y \times d_x d_y B}$ is sparse and can be represented as a concatenation of $B$ diagonal matrices as $\Hmat = \left[ \Dmat_1, \cdots, \Dmat_B \right]$, where $\left\{\Dmat_b = \mbox{diag}\left( \mbox{Vec}(\Cmat_b) \right) \in \mathbb{R}^{d_x d_y \times d_x d_y}\right\}_{b=1}^B$.
Therefore, the compression ratio of SCI is $B$, and the theoretical analysis have been studies in \cite{jalali2019snapshot}.

\subsection{Why Is Fast Adaptation Important?}\label{sec: adaptation}

Recently, deep-learning based SCI reconstruction models have achieved promising results in both fidelity and inference speed \cite{cheng2020birnat, miao2019lambda}.
To be concrete, given the mask matrices $\{\Cmat_b\}_{b=1}^B$ and a training dataset $\mathcal{Q}$ containing $N$ data pairs $\{\Ymat, \Xmat \}$, a deep reconstruction network $f_{\Thetamat}$ is learned by optimizing
\begin{equation}
    \textstyle \min_{\Thetamat} \sum_{\{\Ymat, \Xmat \} \in \mathcal{Q}} d \left( \Xmat, f_{\Thetamat}(\Ymat, \{\Cmat_{b}\}_{b=1}^B) \right),
\end{equation}
where $d(\cdot)$ is a distance metric such as MSE, and $\Thetamat$ denotes all trainable parameters.
For every testing measurement, unlike optimization based methods performing iterative inferences, deep-learning methods~\cite{cheng2020birnat, miao2019lambda, qiao2020deep} directly use the well-learned model $f_{\Thetamat}$ to perform real-time reconstruction.

However, as the system changes, \ie, masks $\{\Cmat_b\}_{b=1}^B$ become $\{\Cmat^{'}_b\}_{b=1}^B$, the network $f_{\Thetamat}$ trained on $\{\Cmat_b\}_{b=1}^B$ {\em does not} work on the new system, as shown in Fig.~\ref{fig1a}. Yet, in real applications, the mask can be changed due to different practical reasons, such as mask drifting or illumination change~\cite{llull2013coded}. 
If $\Cmat^{'}_b$ has the same size as $\Cmat_b$, one still has the opportunity to re-train the model, at the expense of time. Whereas, if $\Cmat^{'}_b$ has larger spatial size than $\Cmat_b$, directly re-training the network often faces enormous challenges in both GPU memory and training time.

For some video SCI systems, \eg, CACTI system \cite{llull2013coded}, the measurement, mask and signal are {\em spatially decoupled}.
This provides a potential solution to the large-scale video reconstruction problem, \ie, decomposing the video into multiple blocks and reconstructing them separately and in parallel. 
For example, as shown in Fig.~\ref{fig1b},  one can decompose a video with $768\times 1792$ pixels into 21 $256 \times 256$ non-overlapping blocks\footnote{The number of blocks increases with the video spatial size.}.
However, with limited computational resources, for existing deep learning based methods, the training time for these 21 models is on the order of months or even years.
Therefore, if we build a model with the fast adaptation property, we can also apply it to realize large-scale video reconstruction via spatial decomposition. That is to say, we train the model only on a small number of sub-masks and then perform fast adaptation to all sub-masks, which will be discussed specifically in Sec. \ref{sec: large-scale}.
In this manner, our model would be more flexible in real applications.

\section{MetaSCI}
MetaSCI considers data from multiple encoding systems as training. During testing, with fast adaptation, the model is flexible to different systems (with different masks) and ready to scale to large data. 
Toward this end, we start by presenting the task definition of MetaSCI.
Based on it, we construct a CNN backbone and introduce meta parameters into the network architecture.
Then, a hybrid learning algorithm is developed to train the network.


\subsection{Task Definition of MetaSCI}
Since the optical encoder in SCI can typically be modeled accurately, it is widely used to train the reconstruction network on simulated data and test on real data \cite{cheng2020birnat}.
In order to consider the adaptive capability into the network learning, we simulate a multi-system training scenario, which can be realized in multiple ways.
For example, we can randomly generate $M$ sets of masks or crop sub-masks from large-scale masks. Each set of masks corresponds to an encoding system.
Without loss of generality, we denote the masks of these $M$ encoding systems as $\{\Cmat^m_{b}\}_{b=1,m=1}^{B,M}$, all of which have the same spatial size.
Correspondingly, for a ground-truth video $\Xmat$, its snapshot measurement in the $m$-th encoding system is denoted as $\Ymat^m$.

For a specific task $m$, the reconstruction network aims to output a recovered video $\widehat{\Xmat}^{m}$ given the inputs $\Ymat^m$ and $\{\Cmat^m_{b}\}_{b=1}^B $.
Jointly considering all these $M$ tasks, our goal is to learn a reconstruction network, whose parameters can be fast adapted to a specific/new task via a few numbers of updates on this task.

\subsection{Fully Convolutional Network Backbone}\label{sec:4.2}
We start by considering the network structure for a single task. Here we take the $m$-th task as an example.

\begin{figure}[!t]
  \centering
  \subfloat[Fully convolutional network backbone]{\includegraphics[width=\linewidth]{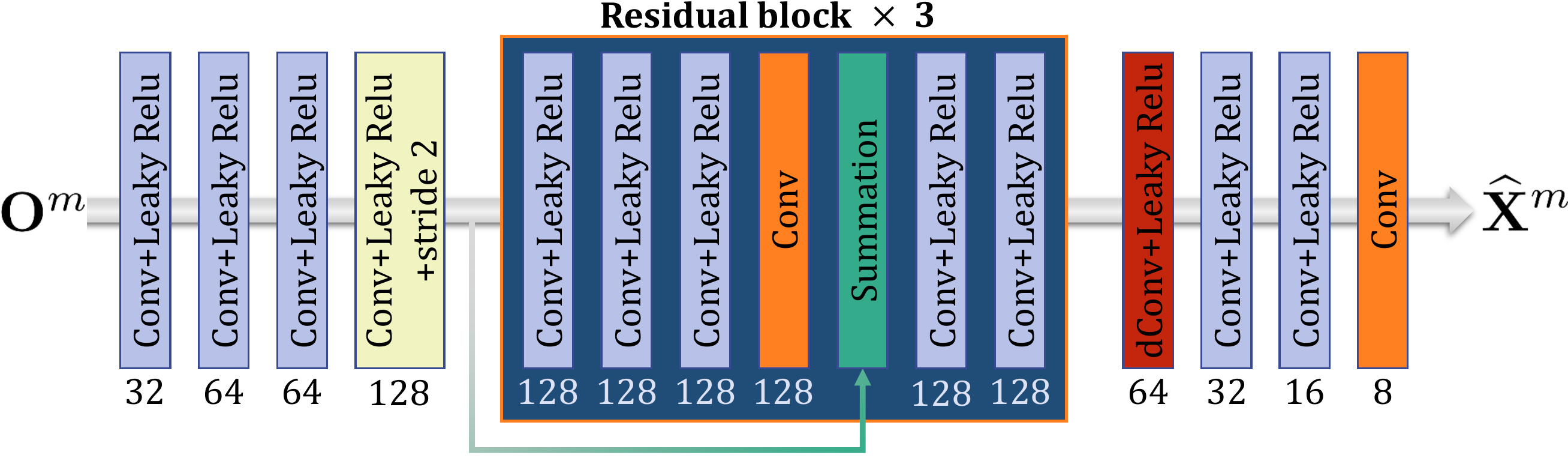}\label{fig3a}} \hfill
  \subfloat[Rank-one convolutional kernel modulation by meta parameters]{\includegraphics[width=\linewidth]{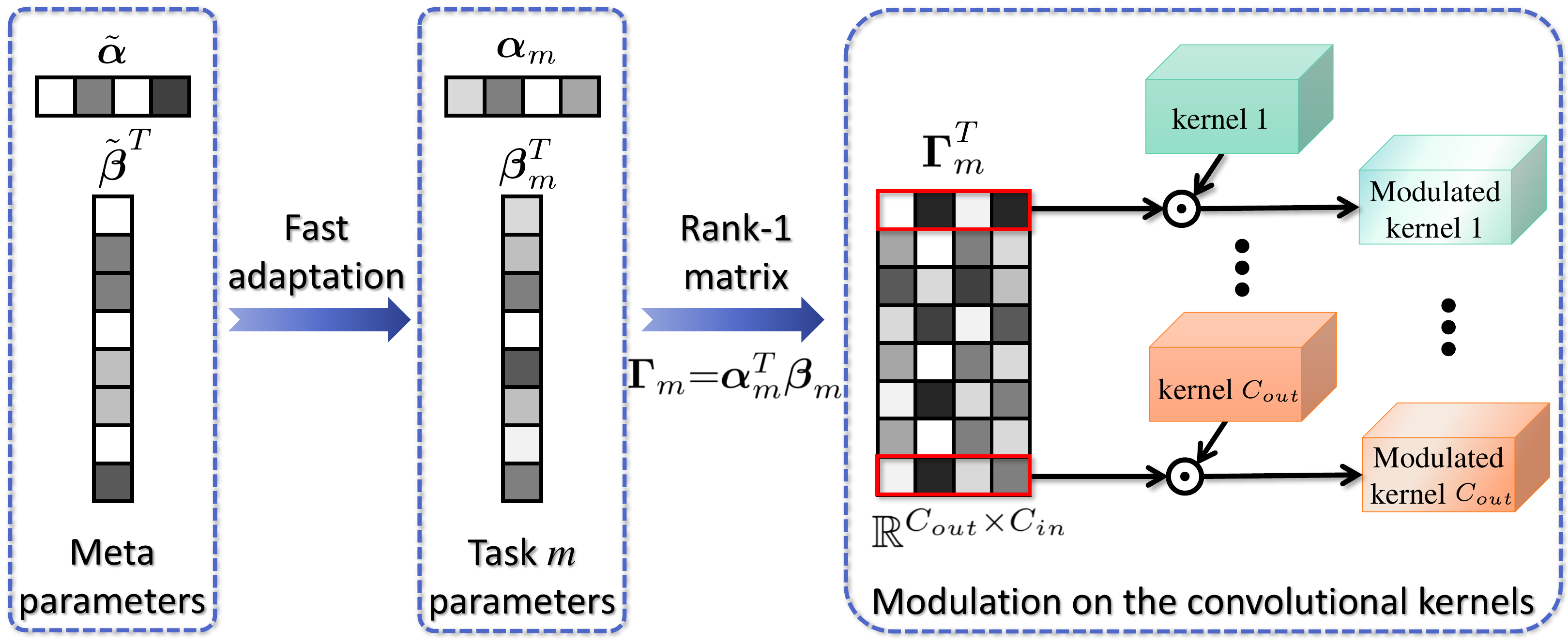}\label{fig3b}} \\
  \vspace{-2mm}
  \caption{(a) The brief structure of our proposed fully CNN backbone for SCI reconstruction, where the numbers denote the number of kernels at each layer. (b) Illustration of our proposed rank-1 convolutional kernel modulation, which is applied to every convolutional layer in (a).}
  \label{fig2: model_2}
   \vspace{-3mm}
\end{figure}

To enhance the motion information in the measurement and achieve better fusion between the measurement and masks, an energy normalization method \cite{cheng2020birnat} is applied, with the normalized measurement $\overline{\Ymat}^m$ derived by 
\vspace{-1mm}
\begin{equation}\label{eq: normalized_measurement}
\textstyle\overline{\Ymat}^m = \Ymat^m \oslash  \left( \sum\nolimits_{b=1}^B \Cmat_{b}^m \right), \vspace{-1mm}
\end{equation}
where $\oslash$ represents the matrix element-wise division.
Considering the fact that $\{\overline{\Ymat}^m \odot  \Cmat_b^m\}_{b=1}^{B}$ can be used to approximate the real modulated frames $\{\Xmat_k \odot  \Cmat_b^m\}_{b=1}^{B}$ \cite{cheng2020birnat}, we fuse all current visual information by
\vspace{-1mm}
\begin{equation} \label{eq: input}
\textstyle \Omat^m = \left[\overline{\Ymat}^m\,,\overline{\Ymat}^m \odot  \Cmat_{1}^m,...,\overline{\Ymat}^m\odot  \Cmat_{B}^m \right]_3, \vspace{-1mm}
\end{equation}
where $\Omat_m \in \mathbb{R}^{d_x \times d_y \times (B+1)}$, and $[\cdot]_3$ denotes concatenation along the third dimension.
Then, we consider $\Omat^m$ as the input of our proposed fully CNN backbone $\mathcal{F}_{\Thetamat_1} (\cdot)$ and achieve reconstruction by
\vspace{-1mm}
\begin{equation} \label{eq: reconstruction}
 \widehat{\Xmat}^{m} = \mathcal{F}_{\Thetamat_1} (\Omat^{m}) \in \mathbb{R}^{d_x \times d_y \times B}, ~ m=1, \cdots, M, \vspace{-1mm}
\end{equation}
where $\Thetamat_1$ denotes the network parameters. 
As shown in Fig.~\ref{fig3a}, the network has three Res-blocks \cite{he2016deep} for better information transmission, each of which has six convolutional layers and one residual connection. 
Detailed network structure is given in the supplementary material (SM).
Different from BIRNAT \cite{cheng2020birnat} that builds a type of CNN only to reconstruct the first frame and employs a time- and memory-consuming RNN to sequentially reconstruct the rest frames, our proposed fully CNN is able to generate all frames quickly in a memory efficient way.

Inspired by the success of meta learning in various fast adaptation tasks \cite{finn2017model,wang2020fusionnet}, we hope to use meta learning to train our proposed backbone, so that it can realize adaptive reconstruction.
A straight forward way is to consider every parameter in $\Thetamat_1$ as a meta-parameter, which is then optimized via a meta-learning algorithm \cite{finn2017model}.
During the test, with the well-learned meta-parameters as initialization, the backbone is adapted to a new system (or task) by a few numbers of updates. 
Though effective, suppose we have $\tilde{M}$ new systems (tasks), the backbone would evolve as $\tilde{M}$ models, that means the volume of parameters would increase $\tilde{M}$ times.
A specific example is discussed in Sec.~\ref{sec: large-scale}. 
This is undesired for the SCI application on edge devices.

To address this challenge, we develop {\em meta-modulation parameters} based on our proposed fully CNN.
Specifically, the following methods are employed to balance the speed and memory.
\begin{list}{\labelitemi}{\leftmargin=10pt \topsep=-1pt \parsep=-1pt}
    \item We assume the backbone is shared by all (training and testing) tasks, and introduce {\em task-specific parameters}, whose volume is much smaller than the backbone, as shown in Fig.~\ref{fig3b}.
    \item  We propose the rank-one convolutional kernel modulation to save the memory during training and also for fast adaptation, with details as follows.
\end{list}

\subsection{Rank-one Kernel Modulation \label{Sec:rank1}}
The basic operation of the backbone in \eqref{eq: reconstruction} is the convolution written as $\Qmat = \Gmat * \Wmat + \Emat$, where $\Gmat \in \mathbb{R}^{I_x \times I_y \times C_{in}}$ is the input feature map, $\Qmat \in \mathbb{R}^{I_x \times I_y \times C_{out}}$ is the output feature map, $\Wmat \in \mathbb{R}^{k_x \times k_y \times C_{in} \times C_{out}}$ is the convolutional kernel, $*$ denotes the convolutional operation, and $\Emat \in \mathbb{R}^{C_{out}}$ is the bias.
For multiple-task purpose, hereby, instead of learning different kernels for different tasks, we utilize a task-specific matrix $\Gammamat_m \in \mathbb{R}^{C_{in} \times C_{out}}$ to modulate the convolutional kernel $\Wmat$.
This is
\begin{equation}\label{eq: modulation}
    \Qmat = \Gmat * (\Wmat \odot \Gammamat_m) + \Emat, \quad m=1, \cdots, M,
\end{equation}
where $\odot$ denotes element-wise product with appropriate broadcasting.
Furthermore, to decrease the volume of the task-specific parameters, we assume $\Gammamat_m$ is a rank-1 matrix, denoted by
\begin{equation}\label{eq: modulation}
   \Gammamat_m = \alphav_m^T \betav_m , \quad  \alphav_m \in \mathbb{R}^{1 \times C_{in}},\quad \betav_m \in \mathbb{R}^{1 \times C_{out}}.
\end{equation}
Compared with $\Gammamat_m$ having $C_{in} \times C_{out}$ parameters, $\alphav_m$ and $ \betav_m$ only have $(C_{in} + C_{out})$ parameters\footnote{For our proposed fully CNN backbone with $B=8$, the total number of meta-modulation parameters is $8.7$k with our developed rank-1 kernel modulation, while it will be $537.8$k with full-matrix modulation.}.

To perform fast adaptation from training to testing tasks, rather than learning $\alphav_m$ and $ \betav_m$ directly, we construct meta-modulation parameters $\Tilde{\alphav} \in \mathbb{R}^{1 \times C_{in}}$ and $\Tilde{\betav}\in \mathbb{R}^{1 \times C_{out}}$ correspondingly, such that $\Tilde{\alphav} $ and $\Tilde{\betav}$ can evolve to $\{\alphav_m\}_{m=1}^{M}$ and $\{\betav_m\}_{m=1}^{M}$, respectively, via a few number of updates.

Denoting all meta-modulation parameters as $\Thetamat_2$, all learnable parameters of MetaSCI are $\Thetamat=\{\Thetamat_{1}, \Thetamat_{2}\}$.
Next, we introduce how to train MetaSCI and then perform fast adaptation.



\subsection{Training and Fast Adaptation \label{Sec:fastAda}}

{\bf{Training:}}
Supposing that each task (corresponding to one set of masks) has $N$ measurements denoted by $\mathcal{T} = \left\{\Ymat_n^m, \{\Xmat_{n,b}^m, \Cmat^m_{b}\}_{b=1}^B\right\}_{n=1,m=1}^{N,M}$ for $M$ tasks. The training objective is to minimize the MSE loss between real and recovered videos:
\begin{equation}\label{eq: loss}
\textstyle   \mathcal{L( \Thetamat;\mathcal{T})} = \sum\nolimits_{m=1,n=1,b=1}^{M,N,B} \left\| \Xmat_{n,b}^m - \widehat{\Xmat}_{n,b}^m \right\|_2.
\end{equation}
Different from general network parameters, $\Thetamat$ contains both shared base parameters $\Thetamat_1$ and meta-modulation parameters $\Thetamat_2$. 
Therefore, the adaptation of $\Thetamat_2$ to task-specific parameters should be considered into the learning algorithm.
Specifically, in each iteration, we sample a mini-batch data $\mathcal{T}_{pre,m}$ for the $m$-th task, and run $U$ (often small, set to $3$ in our experiments) iterations of standard gradient descend to obtain task-specific parameters $\Thetamat_{2,m}'$.
Then, based on another mini-batch data $\mathcal{T}_{obj,m}$ and parameters $\{\Thetamat_1, \Thetamat_{2,m}'\}$, we use Adam \cite{kingma2014adam} to update both $\Thetamat_{1}$ and $\Thetamat_{2}$.
Note that, the task-specified parameters $\Thetamat_{2,m}'$ evolved from $\Thetamat_{2}$ is actually a function w.r.t. $\Thetamat_{2}$.
Thus, rather than updating $\{\Thetamat_1, \{\Thetamat_{2,m}'\}_{m=1}^M\}$, we update $\{\Thetamat_1, \Thetamat_2\}$ during training.
The entire training process is exhibited in Algorithm \ref{alg1}.
 \vspace{-5mm}
 \begin{algorithm}[htbp!]
  \caption{Training algorithm of MetaSCI}
\begin{algorithmic}[1]\label{alg1}
  \REQUIRE Step size $\beta$, number of inner-loop $U$.
  \STATE Randomly initialize $\Thetamat=\{\Thetamat_{1}, \Thetamat_{2}\}$.  
  \WHILE{not done} 
  \FOR{all training tasks $m=1,\cdots,M$} 
  \STATE Sample a mini-batch of data \\$\mathcal{T}_{pre,m} = \{\Ymat_n^m, \{\Xmat_{n,b}^m, \Cmat^m_{b}\}_{b=1}^B\}_{n=1}^{N_1}$.
  \STATE Initialize $\Thetamat_{2,m}' \leftarrow \Thetamat_2$ 
  \FOR{$u=1$ to $U$} 
  \STATE $\mathcal{L}_{1} = \mathcal{L} (\{ \Thetamat_1, \Thetamat_{2,m}'\}; \mathcal{T}_{pre,m})$;
  \STATE $\Thetamat_{2,m}' \leftarrow \Thetamat_{2,m}'  -  \beta \nabla_{\Thetamat_{2,m}'}  \mathcal{L}_1$.
 \ENDFOR 
 \STATE Sample another mini-batch of data $\mathcal{T}_{obj,m}$.
  \ENDFOR
 \STATE Obtain loss:
     $\mathcal{L}_2 = \sum_m  \mathcal{L} ( \{\Thetamat_1, \Thetamat_{2,m}'\}; \mathcal{T}_{obj,m})$.
  \STATE Update all parameters $\Thetamat=\{\Thetamat_1, \Thetamat_2\}$ via\\
      $\Thetamat \leftarrow \Thetamat - Adam[\mathcal{L}_2]$.
  \ENDWHILE
\end{algorithmic}\label{alg1}
\end{algorithm}

\noindent{\bf{Fast adaptation:}}
After training, we obtain the well-learned base parameters $\Thetamat_1$ and meta-modulation parameters $\Thetamat_2$.
During testing, aiming for fast adaptation, we fix $\Thetamat_1$ and only update $\Thetamat_2$ for a new task (with new masks). 

Given $\tilde{M}$ {\em new tasks}, for task $\tilde{m}=1,\dots,\tilde{M}$, the model parameters is represented by $\{ \Thetamat_1, \Thetamat_{2,\tilde{m}} \}$. $\Thetamat_{2,\tilde{m}}$ is firstly initialized by $\Thetamat_2$.
In every iteration, after sampling a mini-batch data $\mathcal{T}_{ad,\tilde{m}}$, we use Adam to update $\Thetamat_{2,\tilde{m}}$. Algorithm \ref{alg2} exhibits how to perform fast adaptation.

Hereby, we describe the difference between our proposed algorithm (Algorithms \ref{alg1} and  \ref{alg2}) and \cite{finn2017model}.
In \cite{finn2017model}, all parameters are regraded as meta parameters, while our propped model exploits a small number of meta parameters to modulate the base backbone.
As a result, as shown in Fig.~\ref{fig: model_1} and Algorithm \ref{alg2}, when performing fast adaptation, all $\tilde{M}$ tasks share a large fixed (thus no need to update) backbone $\Thetamat_1$ but with a small number of task-specific parameters $\{\Thetamat_{2,\tilde{m}}\}_{\tilde{m}=1}^{\tilde{M}}$ that need to be updated from meta parameters $\Thetamat_2$.
Considering the application of MetaSCI for large-scale SCI reconstruction by fast adaptation discussed in the following Sec. \ref{sec: large-scale}, this distinct property makes MetaSCI memory efficient and can be performed in parallel for multiple new tasks.


 \begin{algorithm}[htbp!]
  \caption{Fast adaptation of MetaSCI}
\begin{algorithmic}[1]
  \REQUIRE $\Thetamat=\{\Thetamat_1, \Thetamat_2\}$: the well-learned base and meta-modulation parameters from Algorithm \ref{alg1}.
  \STATE Initialize $\Thetamat_{2,\tilde{m}}$ = $\Thetamat_2$, $\tilde{m}=1, \cdots, \tilde{M}$ .
  \WHILE{not done}
  \FOR{all testing tasks $\tilde{m}=1,\cdots,\tilde{M}$}
  \STATE Sample a mini-batch of data \\$\mathcal{T}_{ad,\tilde{m}} = \{\Ymat_n^{\tilde{m}},   \{\Xmat_{n,b}^{\tilde{m}}, \Cmat_b^{\tilde{m}}\}_{b=1}^B\}_{n=1}^{N_2}$.
  \ENDFOR
  \STATE Obtain loss:
     $\sum_{\tilde{m}} \mathcal{L} ( \{\Thetamat_1, \Thetamat_{2,\tilde{m}}'\}; \mathcal{T}_{ad,\tilde{m}})$ 
  \STATE Update $\{\Thetamat_{2,\tilde{m}}\}_{\tilde{m}=1}^{\tilde{M}}$ via Adam in parallel. 
  \ENDWHILE
\end{algorithmic}\label{alg2}
\end{algorithm}

\subsection{Efficient Reconstruction for Large-scale SCI Using MetaSCI }\label{sec: large-scale}
Existing deep models, such as BIRNAT \cite{cheng2020birnat}, are difficult to handle large-scale videos due to limited GPU memory. 
Interestingly, besides adapting the model for new masks quickly, fast adaptation makes MetaSCI feasible for large-scale SCI reconstruction.
Without loss of generality, in this section, we take videos of size $2048 \times 2048 \times B$ as an example to illustrate the efficient reconstruction using MetaSCI.

Basically, we can spatially decompose a $2048 \times 2048 \times B$ video into 64 non-overlapping $256 \times 256 \times B$ sub-videos\footnote{Actually, in SM, we discuss the effect of overlapping size in recovering large-scale videos, showing that appropriate overlapping will bring better reconstruction, especially at boundaries.}, corresponding to 64 small tasks.
At training stage, we only randomly choose $M$ ($M \ll 64$) sub-videos as training tasks and construct a training set to learn parameters $\Thetamat$ by Algorithm \ref{alg1}.
And then, we use Algorithm \ref{alg2} to perform fast adaptation for the other $\tilde{M}$ ($\tilde{M}+M=64$) sub-videos. 
Finally, after aggregating all $\tilde{M}+M$ sub-videos, we realize large-scale video SCI recovery in an end-to-end manner.

In our experiments, to accelerate the training process, we find $M=4$ is enough to achieve a good $\Thetamat$.
As a result, when adapting it to other $\tilde{M}$ masks, since all tasks share a large fixed backbone with a small number of learnable task-specific modulation parameters, it is efficient to perform parallel fast adaptation on all $\tilde{M}$ masks with a much smaller model\footnote{If we regard all parameters as meta parameters as \cite{finn2017model}, it needs at least (without overlapping) $213.18$M parameters to recover videos with size of $2048 \times 2048 \times 8$. Our proposed MetaSCI only needs $3.89$M parameters.}; please refer to Fig.~\ref{fig1} for an example.

\section{Experiments}
We evaluate the proposed MetaSCI on both simulated data \cite{cheng2020birnat, liu2018rank, ma2019deep} and real data captured by the SCI cameras \cite{qiao2020deep,Sun17OE}.
Considering that most existing methods can only work on small-scale datasets (benchmarks), for comprehensive comparison, we first evaluate MetaSCI on the simulated benchmark datasets, discussing the  reconstruction performance and the adaptation speed to different masks.
Further, we show the appealing results of MetaSCI on large-scale simulated and real datasets,
while most existing methods fail because of the limitations of memory or speed.

\subsection{Implementation Details of MetaSCI}
The high-speed training videos are acquired using the code 
provided by \cite{cheng2020birnat}, containing about 26,000 videos of size $256\times256\times B$ cropped from the \textit{DAVIS2017} dataset \cite{pont2017} with data augmentation.
To simulate a multi-task scenario, we randomly generate four different sets of binary masks of size $256 \times 256$, \ie, $M=4$.
This means that each task has about 26,000 training samples.
For Algorithm~\ref{alg1}, the number of training epochs is set as 100; the number of inner-loop $U$ is set as 3; step size $\beta = 10^{-5}$; we use the default Adam setting \cite{kingma2014adam}.
During the adaptation by Algorithm \ref{alg2}, we only need 4 epochs to achieve good results.
For MetaSCI in reconstructing the large-scale video, we decompose the videos into overlapping $256 \times 256 \times B$ sub-videos using a spatial interval of 128 pixels.


\subsection{Counterparts and Evaluation Metrics}
We compare our method with three representative optimization-based ones, including GAP-TV \cite{yuan2016generalized}, DeSCI \cite{liu2018rank}, and PnP-FFDNet \cite{yuan2020plug}, and two deep-learning based ones, including U-net \cite{qiao2020deep} and BIRNAT \cite{cheng2020birnat}.
Among them, DeSCI and BIRNAT have achieved the SOTA  performance.

Both peak-signal-to-noise ratio (PSNR) and structural similarity (SSIM) \cite{wang2004image} are employed to evaluate the performance. The adaptation and inference speed are assessed by adaptation time and test time, respectively.
The scalability is evaluated on large-scale scenes.

\begin{table*}[htbp!]
  \caption{The results of PSNR in dB (left entry in each cell), SSIM (right entry in each cell), and running time per measurement in seconds on $256\times256\times8$ simulation benchmarks. The results above double lines denote the testing mask appears at training stage while the results under double lines denote the test mask does NOT appear at the training stage. `FT' and `AD' represent `fine-tuning' and `adaptation', respectively. The AD time is determined when training is converged. Note that, with the same AD time as MetaSCI, BIRNAT only achieves about 27dB in average.}
  \vspace{-3mm}
  \centering
  \resizebox{1.0\textwidth}{!}{
  \begin{tabular}{cccccccc>{\columncolor[gray]{.8}[.5\tabcolsep]}cc}
    Algorithm & \texttt{Kobe} & \texttt{Traffic} &\texttt{Runner} &\texttt{Drop} &\texttt{Aerial} &\texttt{Vehicle} &\texttt{Average} & AD Time & Test Time\\ \hline
    GAP-TV &26.45, 0.845 &20.89, 0.715  &28.81, 0.909 &34.74, 0.970 &25.05, 0.828  &24.82, 0.838 &26.79, 0.858 & 0 &4.2\\
    DeSCI  &{33.25}, 0.952  &{28.72}, 0.925  &38.76, 0.969  &{43.22},  {0.993}  &25.33,  0.860  &27.04,  0.909  &32.72, 0.935 & 0 & 6180\\
    PnP-FFDNet & 30.50, 0.926 & 24.18, 0.828 & 32.15, 0.933 & 40.70, 0.989 & 25.27, 0.829 & 25.42, 0.849 & 29.70, 0.892 & 0& 3.0\\ \hline
    U-net & 29.79, 0.807 & 24.62, 0.840 & 34.12, 0.947 & 36.56, 0.949 & 27.18, 0.869 & 26.43, 0.882 & 29.45, 0.882 & 0& 0.0312\\
    BIRNAT &32.71, 0.950 & 29.33, 0.942 &38.70, 0.976 &42.28, 0.992 &28.99, 0.927 &27.84, 0.927 &33.31, 0.951 & 0&0.16\\ \hline
    	\rowcolor{lightgray}
    MetaSCI & 30.12, 0.907 &26.95, 0.888 &37.02, 0.967 &40.61, 0.985 &28.31, 0.904 &27.33, 0.906 & 31.72, 0.926  & 0 & 0.025\\ \hline \hline
    U-net-w/o-FT & 20.13, 0.221 &16.63, 0.165 &23.15, 0.765 &23.02, 0.502 &22.85, 0.527 &20.94, 0.486 & 21.12, 0.443  & 0 & 0.0312\\ 
    U-net-w-FT & 29.81, 0.811 &24.70, 0.843 &34.31, 0.951 &36.51, 0.950 &26.98, 0.860 &26.54, 0.890 & 29.81, 0.884 & 2013 & 0.0312\\ 
    BIRNAT-w/o-FT & 21.45, 0.243 &18.55, 0.186 &26.67, 0.796 &26.12, 0.539 &24.22, 0.559 &22.29, 0.509 & 23.22, 0.387  & 0 & 0.16\\ 
    BIRNAT-w-FT &32.73, 0.952 & 29.30, 0.941 &38.83, 0.975 &42.16, 0.989 &28.93, 0.923 &27.48, 0.907 &33.23, 0.948 & 20376 & 0.16\\ \hline
    \rowcolor{lightgray}
    MetaSCI & 30.10, 0.905 &27.01, 0.891 &37.01, 0.969 &40.52, 0.982 &28.35, 0.904 &27.22, 0.901 & 31.70, 0.925  & 1004 & 0.025\\
    \hline \hline
  \end{tabular}}
  \label{tab: 256}
\end{table*}
\begin{table*}[htbp!]
  \caption{PSNR, SSIM and running time per measurement in seconds on large-scale simulation data with $B$=8. Note that {\em BIRNAT fails} in these large-scale datasets due to high demanding of GPU memory, as well as other deep learning methods.}
  \centering
  \vspace{-3mm}
  \resizebox{1.0\textwidth}{!}{
  \begin{tabular}{ccccccccc}
    Size & Algorithm & \texttt{Beauty} & \texttt{Bosphorus} &\texttt{HoneyBee} &\texttt{Jockey} &\texttt{ShakeNDry} & Average & Test Time \\
    \hline
    \multirow{3}{*}{$512\times512$} & GAP-TV & 32.13, 0.857 & 29.18, 0.934 & 31.40, 0.887 & 31.01, 0.940 & 32.52, 0.882 & 31.25, 0.900 & 44.67 \\
    & PnP-FFDNet & 30.70, 0.855 & 35.36, 0.952 & 31.94, 0.872 & 34.88, 0.955 & 30.72, 0.875 & 32.72, 0.902 & 14.22  \\
   \rowcolor{lightgray}  & MetaSCI & 35.10, 0.901 & 38.37, 0.950 & 34.27, 0.913 & 36.45, 0.962 & 33.16, 0.901 & 35.47, 0.925 & 0.12\\ \hline \hline
    Size & Algorithm & \texttt{Beauty} & \texttt{Jockey} &\texttt{ShakeNDry} &\texttt{ReadyGo} &\texttt{YachtRide} & Average & Test Time \\
    \multirow{3}{*}{$1024\times1024$} & GAP-TV & 33.59, 0.852 & 33.27, 0.971 & 33.86, 0.913 & 27.49, 0.948 & 24.39, 0.937 & 30.52, 0.924 & 178.11 \\
    & PnP-FFDNet & 32.36, 0.857 & 35.25, 0.976 & 32.21, 0.902 & 31.87, 0.965 & 30.77, 0.967 & 32.49, 0.933 & 52.47 \\
   \rowcolor{lightgray} & 
    MetaSCI & 35.23, 0.929 & 37.15, 0.978 & 36.06, 0.939 & 33.34, 0.973 & 32.68, 0.955 & 34.89, 0.955 & 0.59 \\ \hline \hline
    Size & Algorithm & \texttt{City} & \texttt{Kids} &\texttt{Lips} &\texttt{Night} &\texttt{RiverBank} & Average & Test Time \\
    \multirow{3}{*}{$2048\times2048$} & GAP-TV & 21.27, 0.902 & 26.05, 0.956 & 26.46, 0.890 & 26.81, 0.875 & 27.74, 0.848 & 25.67, 0.894 & 764.75\\
    & PnP-FFDNet & 29.31, 0.926 & 30.01, 0.966 & 27.99, 0.902 & 31.18, 0.891 & 30.38, 0.888 & 29.77, 0.915 & 205.62\\
   \rowcolor{lightgray} & 
    MetaSCI & 32.63, 0.930 & 32.31, 0.965 & 30.90, 0.895 & 33.86, 0.893 & 32.77, 0.902 & 32.49, 0.917 & 2.38 \\ \hline \hline
  \end{tabular}}
  \label{Table:high}
   \vspace{-3mm}
\end{table*}


\subsection{Results on Simulated Small-scale Benchmarks}

The widely used six videos \cite{ma2019deep}, including \textit{Kobe}, \textit{Traffic}, \textit{Runner}, \textit{Drop}, \textit{Vehicle}, and \textit{Aerial}, are considered, with spatial size $256 \times 256$. We follow the simulation setup in \cite{cheng2020birnat,liu2018rank, yuan2020plug}, compressing $B$=8 frames into a snapshot measurement.
Besides the standard testing (training and testing have the same set of masks) as usual, we also conduct experiments to evaluate {\em the performance of adaptation to a new set of masks} during the test. 

\begin{figure*}[!th]
    \centering
    \includegraphics[width=1\linewidth]{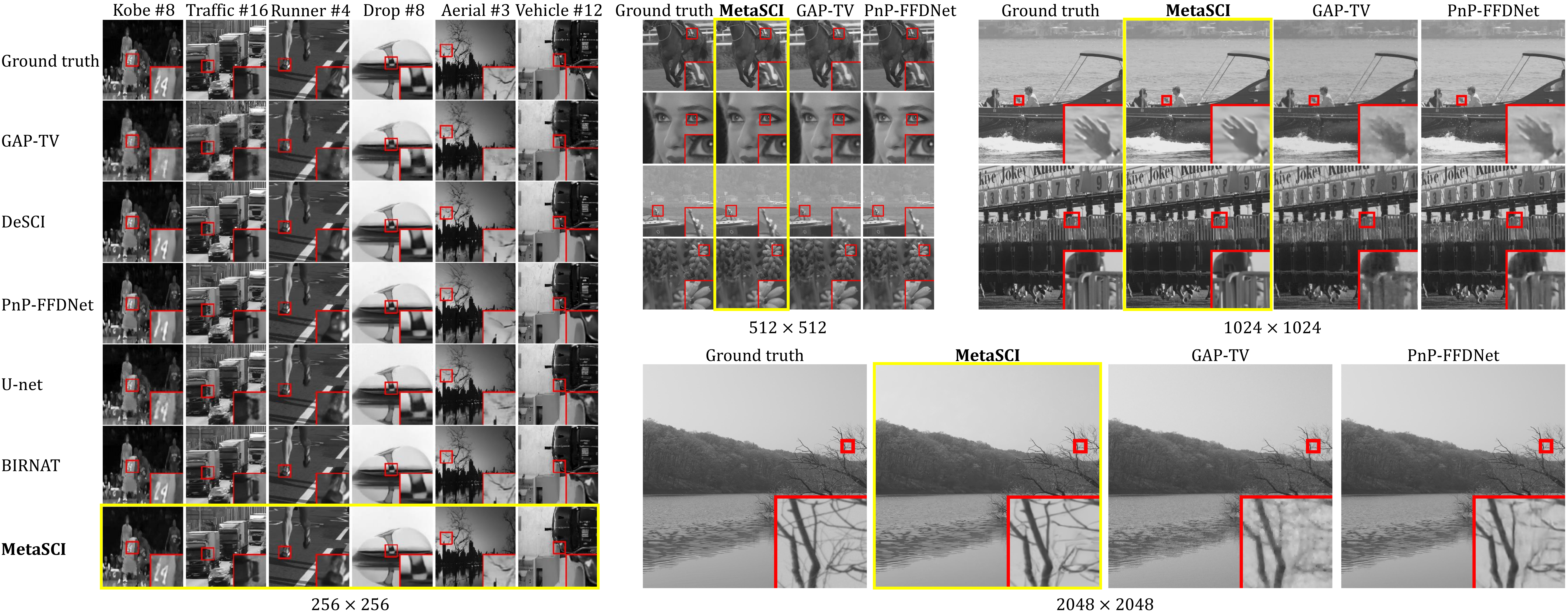}
    \vspace{-6mm}
    \caption{Reconstructed frames on multi-scale simulation benchmarks. 
    }
    \label{fig:high}
    \vspace{-3mm}
\end{figure*}

{\bf{Training and testing have the same masks.}}
The quantitative results on the standard testing are listed in Table \ref{tab: 256} (the upper half).
In general, compared with optimization based ones (in group 1) that need iterative inferences, end-to-end deep learning based methods (in group 2) have much ($\ge$150$\times$) faster testing speed.
Though BIRNAT have achieved the SOTA fidelity performance, the recurrent structure (a sequential forward and then backward frame-by-frame reconstruction) limits its testing speed.
MetaSCI achieves comparable PSNR and SSIM scores with BIRNAT and a superior ($\ge$6$\times$ shorter) testing speed. 
Some reconstructed frames are shown in the left of Fig.~\ref{fig:high}. It can be seen that MetaSCI is able to recover fine details and have little artifacts.

{\bf{Training and testing have different masks.}}
As mentioned in Sec.~\ref{sec: adaptation} and shown in Fig.~\ref{fig1a}, deep-learning based SCI reconstruction models are sensitive to masks.
When the masks change, directly employing the network trained on the original masks to perform testing often results in poor recovery (see U-net-w/o-FT and BIRNAT-w/o-FT in Tab.~\ref{tab: 256}).
Although one can fine-tune the model of U-net or BIRNAT on the testing masks, it is time-consuming.
With half adaptation time, MetaSCI achieves superior fidelity performance than U-net-w-FT.
Using about $1/20$ adaptation time, MetaSCI achieves comparable fidelity performance with BIRNAT-w-FT.

\subsection{Results on Simulated Large-scale Data}
Since limited researches provide large-scale data for SCI, we create a large-scale benchmark, including $512\times512\times8$, $1024\times1024\times8$, and $2048\times2048\times8$ videos cropped from the Ultra Video Group (UVG) dataset \cite{mercat2020uvg}. 
Each scale have five different videos, and the compression ratio is $B=8$. 
Details are provided in the SM.

As the increase of spatial size, existing deep-learning based models often face challenges in both GPU memory and time expense.
Among all existing works, only GAP-TV, PnP-FFDNet, and our proposed MetaSCI can be applied to large-scale SCI reconstruction.
The quantitative comparisons are provided in Tab.~\ref{Table:high}.
Obviously, MetaSCI demonstrates the superior PSNR and SSIM scores, and also testing speed. 
Specifically, MetaSCI outperforms GAP-TV 5.13dB in average, and outperforms PnP-FFDNet 2.62dB in average.
Further, MetaSCI accelerates the test speed over 100$\times$ than GAP-TV and PnP-FFDNet.
As selected reconstructed frames shown in the right of Fig.~\ref{fig:high}, MetaSCI provides sharper boundaries than GAP-TV and less artifacts than PnP-FFDNet.
See recovered videos in the SM.

\subsection{Results on Real Data}
We now apply the proposed MetaSCI to real data. 
Different from simulation, real measurements always have noise inside and thus make the reconstruction more challenging.
In addition to this, the mask is not ideal due to nonuniform illumination and other reasons. 
To handle large scale problems, as mentioned before, for the sake of fast reconstruction, deep learning models need a huge amount of training data and time plus GPU memory. 
Our proposed MetaSCI, on the other hand, provides a solution to train the model on small-scale and efficiently scale to large data as shown in Fig.~\ref{fig1b}. 
We use the real data captured by SCI cameras \cite{qiao2020deep,Sun17OE}, with the mask controlled by a digital micromirror device.
The real data \textit{Domino} and \textit{Water Balloon} are snapshot measurements of size $512 \times 512$, encoding $B=10$ frames. 
The \textit{UCF} snapshot measurement has $850\times1100$ pixels~\cite{Sun17OE}, which is also a compression of 10 frames.

\begin{figure}[htbp!]
    \centering
    \includegraphics[width=\linewidth]{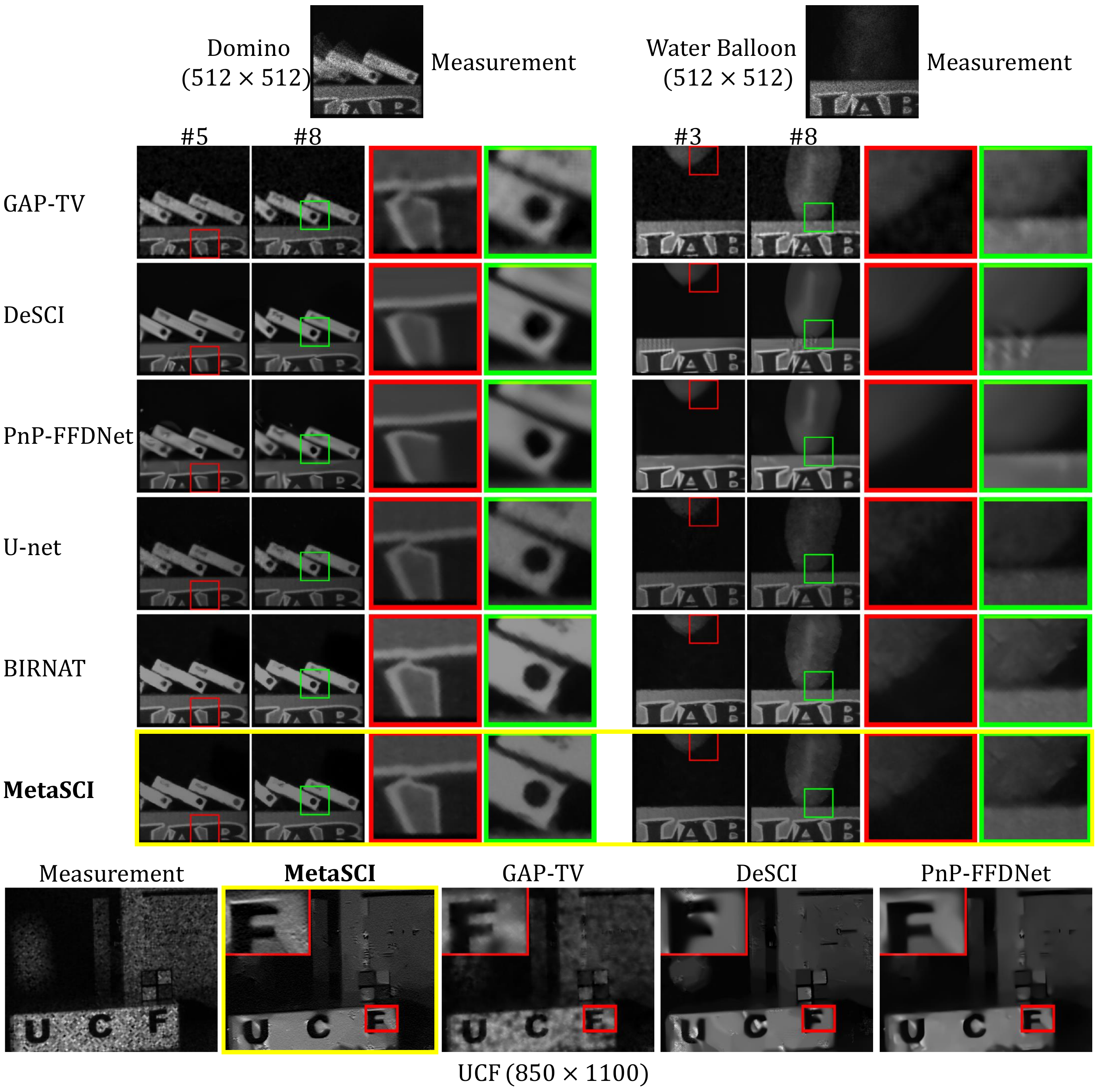}
    \vspace{-6mm}
    \caption{Reconstructed frames on real data, \textit{Domino}, \textit{Water Balloon}, and \textit{UCF}. The test time on the challenging \textit{UCF} is: 0.51s (MetaSCI), 300.84s (GAP-TV), 15045s (DeSCI), 12.52s (PnP-FFDNet).}
    \label{fig: real_data}
    \vspace{-3mm}
\end{figure}

The results compared with other algorithms are shown in 
Fig.~\ref{fig: real_data}, where we can see that the recovered frames of MetaSCI show sharper borders and finer details (recovered videos are in SM).
Meanwhile, on the large scale ($850\times1100$) UCF data, the time cost of MetaSCI is the least after adaptation.
It is worth noting that, in this scale data, BIRNAT fails due to the large memory requirement and we thus only compare with GAP-TV, DeSCI and PnP-FFDnet. DeSCI takes more than 4 hours while MetaSCI only needs 0.51 seconds.


\section{Conclusions and Future Work}
Fast and high quality reconstruction plays a pivot role in inverse problems. This paper takes one step further to devise {\em flexible} reconstruction networks considering the application of video snapshot compressive imaging, where masks can be different for different systems and thus it is desirable to  have fast, accurate and flexible reconstruction algorithms. 
Towards this end, we have developed a meta modulated convolutional network for SCI reconstruction, dubbed MetaSCI, which contains a shared fully CNN as the backbone but being modulated by a small number of meta parameters for different tasks (masks).
In this manner, MetaSCI has accomplished fast adaptation using our proposed algorithms.
MetaSCI is the first end-to-end deep model to realize high-quality and efficient large-scale SCI recovery.

The proposed MetaSCI is actually a backbone-free framework, in which different deep models can be used. 
Furthermore, how to realize fast adaptation among different compression ratios is also a practical problem, which would be considered in the future.

In addition to the video SCI considered in this paper, our proposed MetaSCI can also be used in the spectral SCI reconstruction where large-scale real data are available~\cite{meng2020end,meng2020snapshot,Yuan15_JSTSP_CASSI_SI} and different networks~\cite{Huang2021_CVPR_GSMSCI,miao2019lambda,Zheng20_PRJ_PnP-CASSI} have been proposed for the inversion. MetaSCI is able to speed up the training process and reduce the model size. 


{\small
\bibliographystyle{ieee_fullname}
\bibliography{egbib}
}

\clearpage
\onecolumn
\appendix

\section{Network Structure}

\begin{figure}[!h]
  \centering
  \includegraphics[width=120mm]{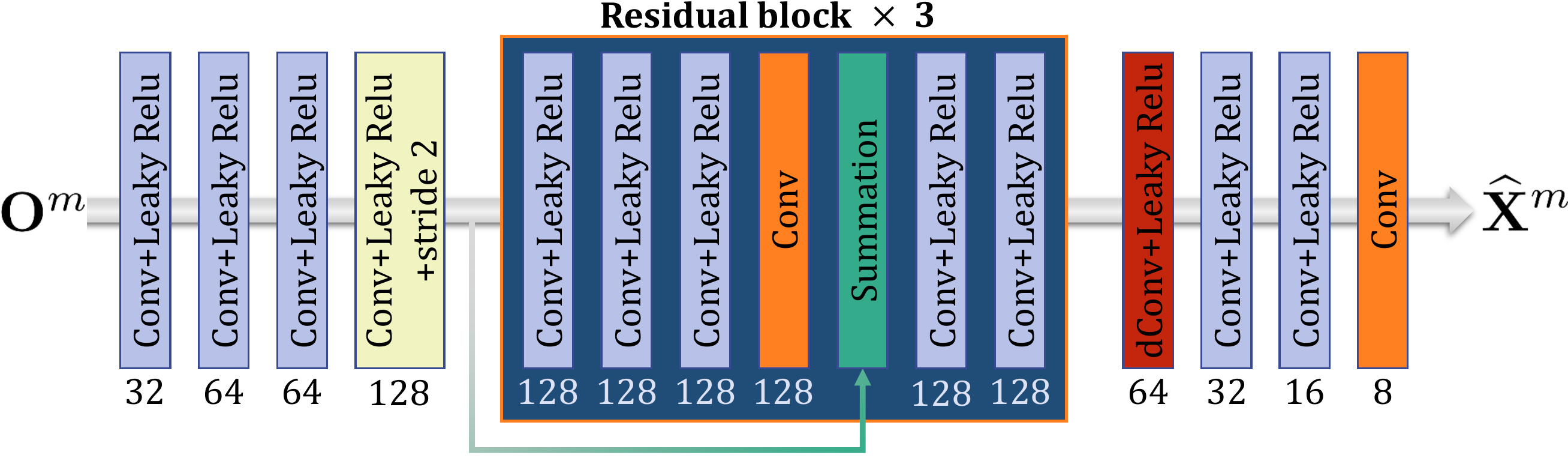}
  \caption{The brief structure of our proposed fully CNN backbone for SCI reconstruction, where the numbers denote the number of kernels at each layer.}\label{fig:appendix1}
\end{figure}

In this section, we provide the detailed network structure of the proposed fully convolutional network backbone in Section \ref{sec:4.2}, with the illustrations shown in Fig.~\ref{fig:appendix1} and Table~\ref{tab4}.

\begin{table}[!h]
  \centering
  \caption{Network Structure of the proposed fully CNN backbone for SCI reconstruction.}\label{tab4}
  \begin{tabular}{c|c|c|c|c}
    \toprule
       & Module  & Stride & Kernel Size & Output Size \\ \midrule
    \multirow{4}{*}{$\times 1$} & Conv. + Leaky Relu & 1 & $5 \times 5 \times (B+1) \times 32$ & $256 \times 256 \times 32$ \\
        & Conv. + Leaky Relu & 1 & $3 \times 3 \times 32 \times 64$ & $256 \times 256 \times 64$ \\
        & Conv. + Leaky Relu & 1 & $1 \times 1 \times 64 \times 64$ & $256 \times 256 \times 64$ \\
        & Conv. + Leaky Relu & 2 & $3 \times 3 \times 64 \times 128$ & $128 \times 128 \times 128$ \\ \midrule
    \multirow{6}{*}{$\times 3$} & Conv. + Leaky Relu & 1 & $3 \times 3 \times 128 \times 128$ & $128 \times 128 \times 128$ \\   
        & Conv. + Leaky Relu & 1 & $1 \times 1 \times 128 \times 128$ & $128 \times 128 \times 128$ \\
        & Conv. + Leaky Relu & 1 & $3 \times 3 \times 128 \times 128$ & $128 \times 128 \times 128$ \\
        & Conv. + Leaky Relu & 1 & $3 \times 3 \times 128 \times 128$ & $128 \times 128 \times 128$ \\
        & Summation & -- & -- & $128 \times 128 \times 128$ \\
        & Conv. + Leaky Relu & 1 & $3 \times 3 \times 128 \times 128$ & $128 \times 128 \times 128$ \\
        & Conv. + Leaky Relu & 1 & $1 \times 1 \times 128 \times 128$ & $128 \times 128 \times 128$ \\ \midrule
    \multirow{4}{*}{$\times 1$} & dConv. + Leaky Relu & 1 & $3 \times 3 \times 64 \times 128$ & $256 \times 256 \times 64$ \\
        & Conv. + Leaky Relu & 1 & $3 \times 3 \times 64 \times 32$ & $256 \times 256 \times 32$ \\
        & Conv. + Leaky Relu & 1 & $1 \times 1 \times 32 \times 16$ & $256 \times 256 \times 16$ \\
        & Conv. + Leaky Relu & 1 & $3 \times 3 \times 16 \times B$ & $256 \times 256 \times B$ \\ \bottomrule
  \end{tabular}

\end{table}

\section{Simulated Large-scale Benchmark}

Since limited researches provide large-scale data for SCI, we create a large-scale benchmark, including $512\times512\times8$, $1024\times1024\times8$, and $2048\times2048\times8$ videos cropped from the Ultra Video Group (UVG) dataset \cite{mercat2020uvg}. 
Each scale have five different video sequences, and the compression ratio is $B=8$. 
We download the original UVG dataset from \url{http://ultravideo.cs.tut.fi/#testsequences}, with the parameters listed in Table~\ref{tab5}.
Since the motion speed varies among different videos, for each video, we sequentially choose 64 frames at an interval of `\textit{Step}' frame(s).
For example, for the video `\textit{Beauty}' of spatial size $512\times512$, we extract 1 frame from every 5 frames in the original video, and totally obtain $64$ frames.
\begin{table}[!t]
    \centering
    \caption{Details of the Simulated Large-scale Benchmark.}
    \resizebox{1.0\textwidth}{!}{
    \begin{tabular}{c|c|c|c|c|c|c|c|c}
        \toprule
        \mr{2}{*}{Spatial Size} & \mr{2}{*}{Name} & \mc{4}{c|}{Original data} & \mc{3}{c}{Processing}   \\
         & & Resolution & Bit depth & Format & Container & Step & Num. of frames & Num. of measurements \\ \midrule
        \mr{5}{*}{$512\times512$} & Beauty & $1920\times1080$ & 8 & YUV & RAW & 5 & 64 & 8 \\
          & Bosphorus & $1920\times1080$ & 8 & YUV & RAW & 5 & 64 & 8 \\
          & HoneyBee & $1920\times1080$ & 8 & YUV & RAW & 1 & 64 & 8 \\
          & Jockey & $1920\times1080$ & 8 & YUV & RAW & 1 & 64 & 8 \\
          & ShakeNDry & $1920\times1080$ & 8 & YUV & RAW & 3 & 64 & 8 \\ \midrule
        \mr{5}{*}{$1024\times1024$} & Beauty & $1920\times1080$ & 8 & YUV & RAW & 5 & 64 & 8 \\
          & Jockey & $1920\times1080$ & 8 & YUV & RAW & 5 & 64 & 8 \\
          & ReadyGo & $1920\times1080$ & 8 & YUV & RAW & 5 & 64 & 8 \\
          & ShakeNDry & $1920\times1080$ & 8 & YUV & RAW & 3 & 64 & 8 \\
          & YachtRide & $1920\times1080$ & 8 & YUV & RAW & 9 & 64 & 8 \\ \midrule
        \mr{5}{*}{$2048\times2048$} & City & $3840\times2160$ & 8 & YUV & RAW & 8 & 64 & 8 \\
          & Kids & $3840\times2160$ & 8 & YUV & RAW & 5 & 64 & 8 \\
          & Lips & $3840\times2160$ & 8 & YUV & RAW & 5 & 64 & 8 \\
          & Night & $3840\times2160$ & 8 & YUV & RAW & 5 & 64 & 8 \\
          & RiverBank & $3840\times2160$ & 8 & YUV & RAW & 5 & 64 & 8 \\ \bottomrule
    \end{tabular}}
    \label{tab5}
\end{table}

\section{Discussion on the Overlapping Size in Large-scale Video Reconstruction}

\begin{table}[!t]
    \centering
    \caption{Number of sub-tasks on large-scale video reconstruction with different overlapping sizes.}
    \begin{tabular}{c|c|c|c}
        \toprule
        \mr{2}{*}{Overlapping size} & \mc{3}{c}{Scale} \\
                & $512\times512$ & $1024\times1024$ & $2048\times2048$  \\ \midrule
         0      & 4          & 16            & 64 \\
         64     & 9          & 25            & 100 \\
         128    & 9          & 49            & 196 \\
         192    & 25         & 100            & 400 \\ \bottomrule
    \end{tabular}
    \label{tab6}
\end{table}

\begin{table}[!t]
    \centering
    \caption{PSNR on large-scale video reconstruction with different overlapping sizes.}
    \begin{tabular}{c|c|c|c}
        \toprule
        \mr{2}{*}{Overlapping size} & \mc{3}{c}{Scale} \\
                & $512\times512$ & $1024\times1024$ & $2048\times2048$  \\ \midrule
         0      & 34.81          & 34.18            & 31.87 \\
         64     & 35.25          & 34.66            & 32.28 \\
         128    & 35.47          & 34.89            & 32.49 \\
         192    & 35.50          & 34.87            & 32.51 \\ \bottomrule
    \end{tabular}
    \label{tab6}
\end{table}

\end{document}